\documentclass[%
 reprint,
 amsmath,amssymb,
 aps,
 prl,
 longbibliography,
 lengthcheck,%
]{revtex4}

\usepackage{graphicx}
\usepackage{dcolumn}
\usepackage{bm}
\usepackage{hyperref}

\usepackage{pstricks,pst-3d,pst-text,pst-node}

\def\nten#1{\mbox{\boldmath{$\rm #1$}}}
\def\nvec#1{\mbox{\boldmath{$\rm #1$}}}

\def\deps{\dot\varepsilon}

\newpsobject{showgrid}{psgrid}{subgriddiv=1,griddots=10,gridlabels=10pt}

\begin{document}

\preprint{APS/123-QED}

\title{Power--Law Creep from Discrete Dislocation Dynamics}

\author{Shyam M. Keralavarma$^1$}

\author{T. Cagin$^{2,3}$}

\author{A. Arsenlis$^4$}

\author{A. Amine Benzerga$^{1,2}$}
\email{benzerga@tamu.edu}

\affiliation{$^1$ Department of Aerospace Engineering, Texas A\&M University, 
             College Station, TX 77843, USA}
\affiliation{$^2$ Materials Science and Engineering Program, Texas A\&M University, 
             College Station, TX 77843, USA}
\affiliation{$^3$ Department of Chemical Engineering, Texas A\&M University, 
             College Station, TX 77843, USA}
\affiliation{$^4$ Lawrence Livermore National Laboratory, Livermore, CA 94551, USA}

\date{\today}

\begin{abstract}
We report two-dimensional
discrete dislocation dynamics simulations of combined dislocation
glide and climb leading to `power-law' creep in a model aluminum crystal.
The approach fully accounts for matter transport due to vacancy diffusion
and its coupling with dislocation motion.
The existence of quasi-equilibrium or jammed states under the applied creep
stresses enables observations of diffusion and climb over time scales
relevant to power-law creep.
The predictions for the creep rates and stress 
exponents fall within experimental ranges, indicating that the underlying
physics is well captured.
\end{abstract}

\maketitle

Dislocations are the main carriers of deformation in crystal plasticity 
\cite{FriedelBook}.
The glide motion of these line defects dominates at low homologous temperatures,
whereas their climb, a nonconservative motion mediated by the absorption or emission
of lattice vacancies, becomes important in high-temperature deformation (creep)
\cite{CaillardBook}.
It is generally believed that the creep strain is mainly produced by dislocation glide
at a rate set by dislocation climb \cite{CaillardBook,Clouet11}.
However, a detailed analysis of the phenomenon is lacking due to the complexity 
of incorporating both vacancies and dislocations in a single computational framework.
As a thermally activated process, the diffusion of vacancies occurs over times scales 
that are much longer than can be accessed by molecular dynamics 
\cite{Gumbsch99,Wang03}, and available dislocation dynamics formulations 
\cite{Bulatov98,Ghoniem00,Bulatov06,Gomez06}
do not account for the nonlinear vacancy--dislocation interactions inherent to climb
\cite{Kabir10,Clouet11}.
Only recently have dislocation glide and climb been simultaneously considered in
the simulation of prismatic loop coarsening \cite{Bako11}.
In this Letter, we extend this approach to simulate power-law creep.
The approach fully utilizes quasi-equilibrium or `jammed' dislocation states
under the low creep stresses to effectively bridge the fine time
scales of dislocation glide with the coarse time scales of diffusion-controlled climb
in a single simulation. It also employs 
a variational principle to derive boundary conditions for the coupled
problem and a dislocation climb model with atomistic
fidelity \cite{Clouet11}.

When the climb motion of noninteracting, pinned edge dislocation segments normal
to the slip plane is assumed to proceed at a velocity proportional to the 
applied stress ($\nten\sigma$) \cite{Mott51,HirthBook}, the steady-state creep
rate exhibits a power-law stress dependence $\deps \propto \sigma^n$ with $n=3$
\cite{Weertman55}. If glide is considered in a creep model, 
Weertman has shown that the stress exponent increases to $n=4.5$; 
see \cite{Weertman75} and references therein.
In fact, the exponent is between 4 and 8 from experiments \cite{Frost82}, which
hints to more complex glide--climb couplings in dislocation creep.
Interestingly, recent atomistic simulations, based on a kinetic Monte Carlo
scheme, and accounting only for climb yielded $n \approx 5$ in bcc iron 
\cite{Kabir10}. However, when this prediction is extrapolated to lower,
realistic dislocation densities and applied stress levels, $n$ is found
to be no more than 3.5 \cite{Clouet11}, consistent with any creep model
based on pure climb \cite{Weertman75}.
Here, we also explore the extent to which the explicit consideration of both climb 
and glide delivers experimentally reported stress exponents.

In the current literature, three-dimensional discrete dislocation dynamics 
formulations of coupled glide and climb remain scarce \cite{Bako11} and do not 
address creep predictions. Some of the remaining challenges involve the disparate time
scales and the precise incorporation of couplings between elasticity and diffusion.
Therefore, the investigation of creep is of considerable importance, but even
in two dimensions (2D), this problem has not been solved yet.

\begin{figure*}[t]
\centering
\includegraphics{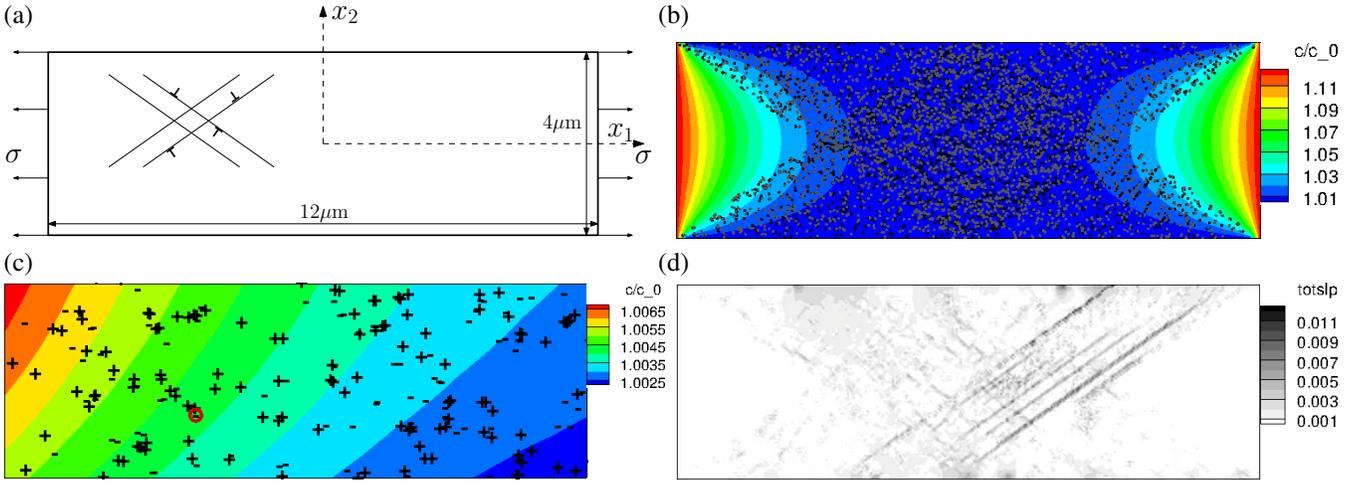}
\caption{(color online) 
(a) Sketch of computational specimen containing discrete edge dislocations
on two independent slip systems oriented at $\pm 35.25^{\circ}$ with respect to
the axis of loading (horizontal).
(b) Contours of vacancy concentration superposed on the positions of dislocations
(positive: black; negative: gray) before the first climb event in a creep 
simulation ($\sigma=40$ MPa, $T=400$ K).
(c) Magnified view of a $1.5 \times 0.5 \mu$m
region around the climbing dislocation (red circle).
(d) Typical contours of slip over a time interval of 200s around $t=4000$s
in the same creep simulation.}
\label{fig:creepbar}
\end{figure*}

We consider a simplified 2D model specimen subjected to plane strain uniaxial
loading at constant average stress with traction-free top and bottom surfaces,
Fig.~\ref{fig:creepbar}(a).
The specimen initially contains discrete edge dislocations, dislocation sources 
and obstacles embedded in a linear elastic material.
The instantaneous state of the system is characterized by the positions
of all dislocations, $\nvec x^i(t)$ ($i=1,2,3..)$, and a continuous
field of the fractional vacancy concentration, $c(\nvec x,t)$.
The long- and short-range dislocation interactions are handled as in 
\cite{Giessen95,Benzerga04MSMSE} using the finite-element method.
The driving force for
dislocation motion is the generalized Peach--Koehler force while the driving force for
vacancy diffusion is the gradient of the chemical potential, $\mu$, both of which may be
obtained from derivatives of the Gibbs free energy function $G(\nvec x^i,c)$.
The governing equations for vacancy diffusion are
\begin{equation}
\dot c = -\nabla\cdot\nvec J + \dot c_{\rm src}
\label{eq:continuity}
\end{equation}
\begin{equation}
\nvec J = -\frac{D\Omega}{kT}\nabla\mu
\label{eq:fick}
\end{equation}
\begin{equation}
\mu = \frac{kT}{\Omega} \left[ \frac{E_f}{kT} -
\frac{p\Omega_v}{kT} + \log{\frac{c}{(1-c)}} \right]
\label{eq:mudef}
\end{equation}
Equation~(\ref{eq:continuity}) follows from mass conservation
where $\nvec J$ denotes the volumetric flux of vacancies and
$\dot c_{\rm src}$ is a production term (see below).
In~(\ref{eq:fick}) and (\ref{eq:mudef}), 
$k$ is Boltzmann's constant, $T$ the absolute temperature, 
$\Omega$ the atomic volume, $E_f$ the vacancy formation energy, 
$p = p(\nvec x, t)$ denotes the hydrostatic pressure field, 
$\Omega_v$ the vacancy relaxation volume, and
$D = D_0\exp{\left( -\frac{E_m}{kT} \right)}$ the vacancy diffusion coefficient, 
with $E_m$ the vacancy migration energy. In the absence of pressure gradients,
equation~(\ref{eq:fick}) reduces to Fick's first law of diffusion.
The glide velocity, $v_g^i$, of
dislocation $i$ is taken to be proportional to the glide component of the
Peach-Koehler force, $f_g^i$, as
\begin{equation}
v_g^i = f_g^i/B(T)
\label{eq:glvel}
\end{equation}
where the drag factor $B$ varies linearly with temperature \cite{HirthBook}.
In the most fundamental formulation, the climb velocity, $v_c^i$,
is determined by mass conservation from
$b^i v_c^i = \int_{\partial C^i} \nvec J \cdot \nvec n \,\,{\rm d}S$
where $\partial C^i$ denotes the dislocation core boundary with
inward unit normal $\nvec n$, so that $\dot c_{\rm src}=0$ in~(\ref{eq:continuity}).
To circumvent the computational complexity of such an approach,
$v_c^i$, is estimated
from the {\it net} flux of vacancies to/from the dislocation core and mass conservation,
over a larger volume
under steady state climb conditions and assumed radial symmetry around the core
\cite{HirthBook,Mordehai08}
\begin{equation}
v_c^i = -\eta\frac{D}{b^i}\left[ c_0\exp{\left(-\frac{f_c^i\Omega}{b^ikT}\right)} - c \right]
\label{eq:clvel}
\end{equation}
The first term in brackets is the concentration of vacancies in equilibrium
with the core,
where $f_c^i$ denotes the climb component of the Peach-Koehler force,
$b^i$ is the magnitude of the Burgers vector and $c_0=\exp{({-E_f}/{kT})}$ is the
equilibrium vacancy concentration in a bulk material at temperature $T$.
$c$ is the ambient vacancy concentration away from the dislocation core, here obtained
from the solution of the global diffusion equations interpolated to the dislocation position
under the assumption that the dislocation core radius is much smaller
than the length scale of the gradients of $c$.
$\eta$ is a constant of order unity.
As shown in \cite{Clouet11} the performance of analytical estimate~(\ref{eq:clvel})
against atomistic simulations is remarkable.
Consistent with this, the term $\dot c_{\rm src}$ is added to~(\ref{eq:continuity})
to account for the net absorption/emission of vacancies in the volume element.

In order to perform creep simulations over time durations sufficient to establish a
steady state, we used an adaptive scheme to increment the simulation time step
proceeding as follows.
We first relaxed the initial dislocation microstructures at zero stress until the
dislocations attained quasi-equilibrium positions. 
Thereafter, we applied the creep stress (below the nominal yield stress of the specimen) 
and performed the simulation using a small time step of 0.5 ns to resolve glide-related
events until the overall strain attained a constant value as determined by measuring
the average slope of the strain vs. time plot over a predefined interval.
Attainment of such a `jammed' or quasi-equilibrium state enables observation of creep
deformation over macroscopically relevant time scales.
Indeed, at that point, we computed the evolution of the vacancy field by solving equations
(\ref{eq:continuity})--(\ref{eq:mudef}) 
using the finite element method and a much larger value of the time step, 
dependent on temperature as per the analytical estimate~(\ref{eq:clvel}) for climbing 
to a neighboring slip plane.
Dirichlet boundary conditions were imposed for $c$ corresponding to the equilibrium
concentrations at the boundaries consistent with the imposed tractions. The initial $c$
field was specified according to the steady state solution of equation (\ref{eq:continuity}),
with $\dot c = 0$ and $\dot c_{\rm src} = 0$.
(Due to the much larger value of the time step used, the diffusion
equations are solved using a fully implicit algorithm as opposed to a simple forward Euler
scheme for the glide steps.)
When the first `activation event' is detected, i.e. when any
dislocation climbs to a new slip plane thereby enabling further glide, the time step
is reverted to the fine 0.5 ns value.
Note that the dislocation dynamics (DD) and the vacancy diffusion problems are inherently
coupled since the frequency and locations of the dislocation climb events are determined
by the vacancy distribution according to equation (\ref{eq:clvel}) and any
production/annihilation of vacancies due to climb affects the evolution of the
vacancy field through the production term in equation (\ref{eq:continuity}).

We performed all simulations using physical properties of fcc aluminum:
Young's modulus $E=70$ GPa, Poisson's ratio $\nu=0.33$,
$E_f=0.67$ eV, $E_m=0.61$ eV,
$D_0=1.51\times10^{-5}$ m$^2$/s \cite{Freund02}.
Also, $B(T)=10^{-4}\times(T/300)$ Pa~s, $\Omega=16.3$ \AA{}$^3$ and $b=0.25$ nm.
The relaxation volume of a vacancy in Al is neglected and the geometry factor $\eta$
in~(\ref{eq:clvel}) is taken to be unity. 
The temperature dependence of the elastic properties is neglected. 
Fig.~\ref{fig:creepbar}(b) shows the dislocation
positions before the first climb event in a creep simulation
of a $12\times4$ $\mu$m$^2$ specimen at $T=400$ K and $\sigma=40$ MPa
superposed with contours of the normalized vacancy concentration $c/c_0$.
Fig.~\ref{fig:creepbar}(c) shows a magnified view of a $1.5\times0.5$ $\mu$m$^2$ region
around the climbing dislocation.
Fig.~\ref{fig:creepbar}(d) shows the contours of cumulated plastic slip over a 200 seconds
time interval in the same specimen. Unlike in a plasticity simulation by pure glide,
which shows sharp slip traces oriented along the slip planes, Fig.~\ref{fig:creepbar}(d)
shows some bands oriented normal to the slip planes indicating dislocation climb activity.

\begin{figure}
\centering
\includegraphics{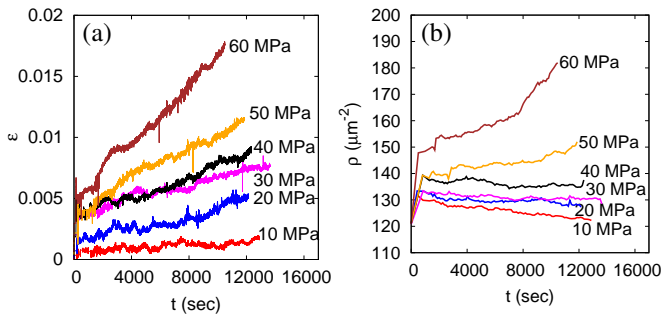}
\caption{(a) Creep curves showing the total strain, $\epsilon$, as a function of time, $t$,
for various values of the creep stress at $T=400$ K. (b) Corresponding evolution of the
dislocation density, $\rho$.}
\label{fig:st}
\end{figure}

\begin{figure}[t]
\centering
\includegraphics[width=0.9\columnwidth]{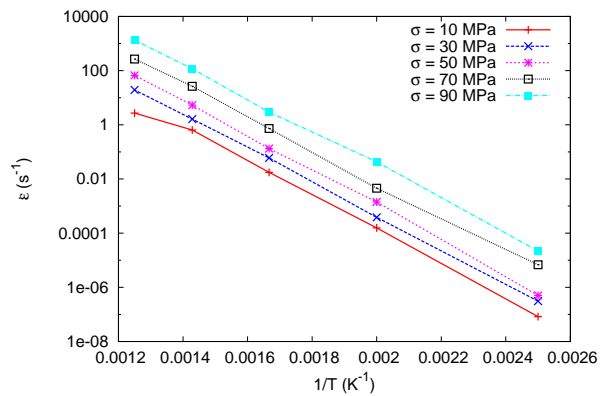}
\caption{Variation of the creep strain rate as a function of reciprocal temperature for various
values of the creep stress. The Arrhenius activation energy for creep estimated from the
average slopes of the above plots is approximately 120 kJ/mol.}
\label{fig:activation}
\end{figure}

A large number of creep simulations have been carried out
in the temperature range $T$ = 400--800 K
($0.43T_m$--$0.86T_m$, $T_m=933$ K for Al) and stresses $\sigma=10$--$90$ MPa
(spanning the range $10^{-4}G$--$10^{-3}G$ for the resolved shear
stress with $G=E/2/(1+\nu)$ the shear modulus).
Fig.~\ref{fig:st}(a) shows the creep curves obtained from the DD simulations at 
$T=400$ K ($0.43T_m$) and several values of the creep stress.
The high temperature DD simulations yield a steady
state creep response with the strain rate increasing with the applied stress.
A characteristic feature of steady state creep is that the material microstructure
remains unchanged with time on average. An average description of the microstructure
in the present problem is the dislocation density, which is plotted as a function of time
in Fig.~\ref{fig:st}(b) corresponding to the creep curves in Fig.~\ref{fig:st}(a).
Following a rapid initial transient, the dislocation density evolves
slowly with time except at high stresses indicating that steady state conditions have
been attained. The steady state creep rates depend exponentially on the temperature and the
relationship between the creep rates and the temperature follows an Arrhenius type equation
\begin{equation}
\dot\epsilon = \dot\epsilon_0\exp{\left(-\frac{Q}{kT}\right)}
\label{eq:creepact}
\end{equation}
where the activation energy for creep, $Q$, is experimentally known to be close to the
activation energy for self diffusion, $E_s=E_f+E_m$. The former may be determined
from the negative slope of the logarithm of the strain rate plotted as a function of
the reciprocal temperature, as shown in Fig.~\ref{fig:activation} for various values
of the creep stress.
Notice that a more or less constant slope is obtained for the activation plot
irrespective of the creep stress and the measured value of $Q=120$ kJ/mol compares
favorably with the value of $E_s=123$ kJ/mol assumed in the simulations.
The emergence of creep from DD simulations (Fig.~\ref{fig:st}(a) and 
Fig.~\ref{fig:activation}) is the main result of this letter.

\begin{figure}[t]
\centering
\includegraphics[width=0.9\columnwidth]{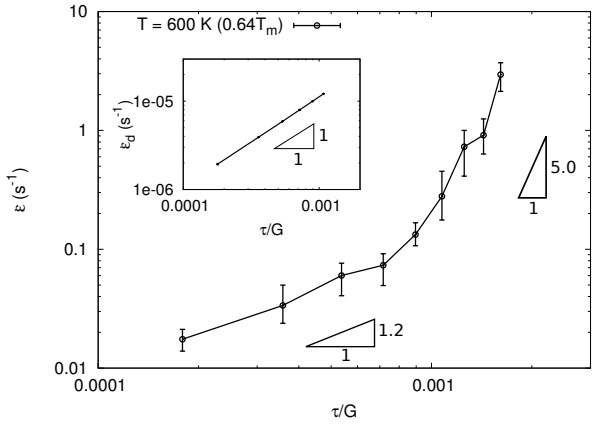}
\caption{Scaling of the net creep strain rate, $\dot\epsilon$, with
the normalized resolved shear stress,
$\tau/G$, at $T=600$ K. The strain rates are averaged over at least three
realizations of the initial microstructure, consisting of randomly distributed dislocations,
point sources and obstacles with a specified average density. The creep exponent, $n$,
is measured as the slope of the curve on the log--log plot.
The inset shows the scaling of the creep rate due to vacancy diffusion 
(Nabarro-Herring creep).}
\label{fig:scaling}
\end{figure}

The stress dependence of the creep rate has been probed in the temperature range
$T=400$--$800$ K. Fig.~\ref{fig:scaling} plots the steady state creep rates (determined
approximately by a linear least squares fit to the $\epsilon-t$ plots) as a function of
the creep stress at $T=600$ K on a log-log scale and illustrates the typical emergent
behavior.
A minimum of three sets of simulations have been performed
using different realizations of the initial dislocation, source and obstacle structure
for a given value of the creep stress.
These attributes of the initial microstructure are chosen so that the
nominal yield stress in a displacement driven simulation is $\approx 80--90$~MPa.
The slope of the log-log plot gives the stress exponent $n$. The results
show two distinct regimes where the stress exponent approaches unity towards low
values of the creep stress ($\tau\sim10^{-4}G$) while $n\approx5$ is obtained
at high stresses ($\tau\sim10^{-3}G$). 

\begin{figure}[t]
\centering
\includegraphics[width=0.8\columnwidth]{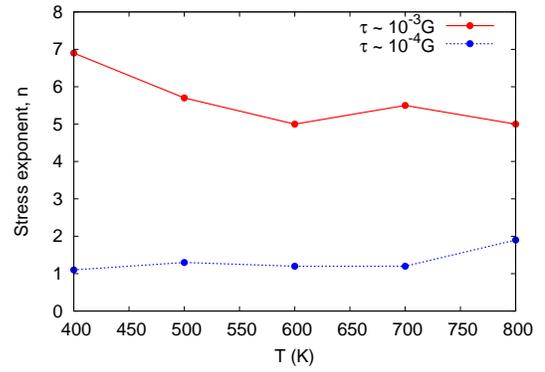}
\caption{Variation of the stress exponent for creep determined from the
simulations as a function of temperature for low ($\sim 10^{-4}G$) and high
($\sim 10^{-3}G$) values of the creep stress.}
\label{fig:exponent}
\end{figure}

The creep strain in Fig.~\ref{fig:st}(a) has two components:
one is mechanical that results from dislocation motion, the other is
diffusive due to transport of matter towards the loaded ends of the specimen. The average
creep rate due to mass transport, $\dot\epsilon_d$, is estimated as the total
volumetric flux of vacancies from the end faces normalized by the volume of the specimen.
The inset in Fig.~\ref{fig:scaling} plots
just the diffusion component of the creep strain rate $\dot\epsilon_d$ as a function
of the stress. The latter yields a stress scaling exponent $n=1$ as expected for the
Nabarro--Herring creep mechanism. Also, it is clear that at $T=600$ K, the diffusion
contribution to the overall creep rate is negligible compared to that due to combined
dislocation glide and climb. The same qualitative behavior as in Fig.~\ref{fig:scaling}
is obtained at all temperatures above $0.4T_m$. Fig.~\ref{fig:exponent}
summarizes the results of our simulations and shows the measured value of $n$
at low and high stresses as a function of temperature.
Values of $n$ close to unity are obtained at low stresses while at high stresses the
predicted values of $n$ span the range $5$--$7$, well within the experimental range of
$4$--$8$. 

The good agreement between the calculated and measured power-law creep 
exponents is a major result of this letter.
We note, however, that a 2D treatment of dislocation dynamics sets
restrictions on (i) the actual degrees of freedom that flexible dislocations
have; and (ii) the incorporation of other recovery mechanisms, such as cross-slip
\cite{Poirier76},
or bypassing versus shearing of precipitates. It remains to be seen what stress
exponents will emerge from fully 3D simulations, properly extended as done here in 2D.

Our results show that meso-scale simulation methods such as DD
can provide new insights into aspects of material behavior heretofore not explained using
continuum approaches. Unlike fully discrete methods such as molecular dynamics, meso-scale
methods offer better scalability to larger and more complex problems. 
Beyond the athermal interactions of
dislocations considered in most DD studies, thermally activated mechanisms of deformation
predominate under different regimes of stress and temperature, as illustrated succinctly
in deformation mechanism maps \cite{Frost82}. Our results
show that power-law exponents of 5 and higher emerge, at realistic levels
of stress and dislocation density, when both climb and glide are modeled
and that the collective behavior of dislocations, which is not amenable to
simple analytical treatment, contributes to the power-law dependence.

We gratefully acknowledge financial support from the
Lawrence Livermore National Security, LLC under Master Task
Agreement No. B575363, LLNL under Contract DE-AC52-07NA27344
and the National Science Foundation 
under grant CMMI-0748187 (A.A.B.).


\begin{thebibliography}{21}
\expandafter\ifx\csname natexlab\endcsname\relax\def\natexlab#1{#1}\fi
\expandafter\ifx\csname bibnamefont\endcsname\relax
  \def\bibnamefont#1{#1}\fi
\expandafter\ifx\csname bibfnamefont\endcsname\relax
  \def\bibfnamefont#1{#1}\fi
\expandafter\ifx\csname citenamefont\endcsname\relax
  \def\citenamefont#1{#1}\fi
\expandafter\ifx\csname url\endcsname\relax
  \def\url#1{\texttt{#1}}\fi
\expandafter\ifx\csname urlprefix\endcsname\relax\def\urlprefix{URL }\fi
\providecommand{\bibinfo}[2]{#2}
\providecommand{\eprint}[2][]{\url{#2}}

\bibitem[{\citenamefont{Friedel}(1964)}]{FriedelBook}
\bibinfo{author}{\bibfnamefont{J.}~\bibnamefont{Friedel}},
  \emph{\bibinfo{title}{{Dislocations}}} (\bibinfo{publisher}{Pergamon Press},
  \bibinfo{address}{Oxford}, \bibinfo{year}{1964}).

\bibitem[{\citenamefont{Caillard and Martin}(2003)}]{CaillardBook}
\bibinfo{author}{\bibfnamefont{D.}~\bibnamefont{Caillard}} \bibnamefont{and}
  \bibinfo{author}{\bibfnamefont{J.}~\bibnamefont{Martin}},
  \emph{\bibinfo{title}{Thermally Activated Mechanisms in Crystal Plasticity}},
  vol.~\bibinfo{volume}{8} (\bibinfo{publisher}{Elsevier Science},
  \bibinfo{address}{Amsterdam}, \bibinfo{year}{2003}).

\bibitem[{\citenamefont{Clouet}(2011)}]{Clouet11}
\bibinfo{author}{\bibfnamefont{E.}~\bibnamefont{Clouet}},
  \bibinfo{journal}{Phys. Rev. B} \textbf{\bibinfo{volume}{84}},
  \bibinfo{pages}{092106} (\bibinfo{year}{2011}).

\bibitem[{\citenamefont{Gumbsch and Gao}(1999)}]{Gumbsch99}
\bibinfo{author}{\bibfnamefont{P.}~\bibnamefont{Gumbsch}} \bibnamefont{and}
  \bibinfo{author}{\bibfnamefont{H.~J.} \bibnamefont{Gao}},
  \bibinfo{journal}{Science} \textbf{\bibinfo{volume}{283}},
  \bibinfo{pages}{965} (\bibinfo{year}{1999}).

\bibitem[{\citenamefont{Wang et~al.}(2003)\citenamefont{Wang, Strachan, Cagin,
  and Goddard}}]{Wang03}
\bibinfo{author}{\bibfnamefont{G.~F.} \bibnamefont{Wang}},
  \bibinfo{author}{\bibfnamefont{A.}~\bibnamefont{Strachan}},
  \bibinfo{author}{\bibfnamefont{T.}~\bibnamefont{Cagin}}, \bibnamefont{and}
  \bibinfo{author}{\bibfnamefont{W.~A.} \bibnamefont{Goddard}},
  \bibinfo{journal}{Phys. Rev. B} \textbf{\bibinfo{volume}{68}},
  \bibinfo{pages}{224101} (\bibinfo{year}{2003}).

\bibitem[{\citenamefont{Bulatov et~al.}(1998)\citenamefont{Bulatov, Abraham,
  Kubin, Devincre, and Yip}}]{Bulatov98}
\bibinfo{author}{\bibfnamefont{V.}~\bibnamefont{Bulatov}},
  \bibinfo{author}{\bibfnamefont{F.~F.} \bibnamefont{Abraham}},
  \bibinfo{author}{\bibfnamefont{L.}~\bibnamefont{Kubin}},
  \bibinfo{author}{\bibfnamefont{B.}~\bibnamefont{Devincre}}, \bibnamefont{and}
  \bibinfo{author}{\bibfnamefont{S.}~\bibnamefont{Yip}},
  \bibinfo{journal}{Nature} \textbf{\bibinfo{volume}{391}},
  \bibinfo{pages}{669} (\bibinfo{year}{1998}).

\bibitem[{\citenamefont{Ghoniem et~al.}(2000)\citenamefont{Ghoniem, Tong, and
  Sun}}]{Ghoniem00}
\bibinfo{author}{\bibfnamefont{N.~M.} \bibnamefont{Ghoniem}},
  \bibinfo{author}{\bibfnamefont{S.~H.} \bibnamefont{Tong}}, \bibnamefont{and}
  \bibinfo{author}{\bibfnamefont{L.~Z.} \bibnamefont{Sun}},
  \bibinfo{journal}{Phys. Rev. B} \textbf{\bibinfo{volume}{61}},
  \bibinfo{pages}{913} (\bibinfo{year}{2000}).

\bibitem[{\citenamefont{Bulatov et~al.}(2006)\citenamefont{Bulatov, Hsiung,
  Tang, Arsenlis, Bartelt, Cai, Florando, Hiratani, Rhee, Hommes
  et~al.}}]{Bulatov06}
\bibinfo{author}{\bibfnamefont{V.~V.} \bibnamefont{Bulatov}},
  \bibinfo{author}{\bibfnamefont{L.~L.} \bibnamefont{Hsiung}},
  \bibinfo{author}{\bibfnamefont{M.}~\bibnamefont{Tang}},
  \bibinfo{author}{\bibfnamefont{A.}~\bibnamefont{Arsenlis}},
  \bibinfo{author}{\bibfnamefont{M.~C.} \bibnamefont{Bartelt}},
  \bibinfo{author}{\bibfnamefont{W.}~\bibnamefont{Cai}},
  \bibinfo{author}{\bibfnamefont{J.~N.} \bibnamefont{Florando}},
  \bibinfo{author}{\bibfnamefont{M.}~\bibnamefont{Hiratani}},
  \bibinfo{author}{\bibfnamefont{M.}~\bibnamefont{Rhee}},
  \bibinfo{author}{\bibfnamefont{G.}~\bibnamefont{Hommes}},
  \bibnamefont{et~al.}, \bibinfo{journal}{Nature}
  \textbf{\bibinfo{volume}{440}}, \bibinfo{pages}{1174} (\bibinfo{year}{2006}).

\bibitem[{\citenamefont{G{\'o}mez-Garcia
  et~al.}(2006)\citenamefont{G{\'o}mez-Garcia, Devincre, and Kubin}}]{Gomez06}
\bibinfo{author}{\bibfnamefont{D.}~\bibnamefont{G{\'o}mez-Garcia}},
  \bibinfo{author}{\bibfnamefont{B.}~\bibnamefont{Devincre}}, \bibnamefont{and}
  \bibinfo{author}{\bibfnamefont{L.~P.} \bibnamefont{Kubin}},
  \bibinfo{journal}{Phys. Rev. Lett.} \textbf{\bibinfo{volume}{96}},
  \bibinfo{pages}{125503} (\bibinfo{year}{2006}).

\bibitem[{\citenamefont{Kabir et~al.}(2010)\citenamefont{Kabir, Lau, Rodney,
  Yip, and Van~Vliet}}]{Kabir10}
\bibinfo{author}{\bibfnamefont{M.}~\bibnamefont{Kabir}},
  \bibinfo{author}{\bibfnamefont{T.~T.} \bibnamefont{Lau}},
  \bibinfo{author}{\bibfnamefont{D.}~\bibnamefont{Rodney}},
  \bibinfo{author}{\bibfnamefont{S.}~\bibnamefont{Yip}}, \bibnamefont{and}
  \bibinfo{author}{\bibfnamefont{K.~J.} \bibnamefont{Van~Vliet}},
  \bibinfo{journal}{Phys. Rev. Lett.} \textbf{\bibinfo{volume}{105}},
  \bibinfo{pages}{095501} (\bibinfo{year}{2010}).

\bibitem[{\citenamefont{Bak{\'o} et~al.}(2011)\citenamefont{Bak{\'o}, Clouet,
  Dupuy, and Bl{\'e}try}}]{Bako11}
\bibinfo{author}{\bibfnamefont{B.}~\bibnamefont{Bak{\'o}}},
  \bibinfo{author}{\bibfnamefont{E.}~\bibnamefont{Clouet}},
  \bibinfo{author}{\bibfnamefont{L.~M.} \bibnamefont{Dupuy}}, \bibnamefont{and}
  \bibinfo{author}{\bibfnamefont{M.}~\bibnamefont{Bl{\'e}try}},
  \bibinfo{journal}{Philosophical Magazine} \textbf{\bibinfo{volume}{91}},
  \bibinfo{pages}{3173} (\bibinfo{year}{2011}).

\bibitem[{\citenamefont{Mott}(1951)}]{Mott51}
\bibinfo{author}{\bibfnamefont{N.~F.} \bibnamefont{Mott}},
  \bibinfo{journal}{Proceedings of the Physical Society. Section B}
  \textbf{\bibinfo{volume}{64}}, \bibinfo{pages}{729} (\bibinfo{year}{1951}).

\bibitem[{\citenamefont{Hirth and Lothe}(1968)}]{HirthBook}
\bibinfo{author}{\bibfnamefont{J.~P.} \bibnamefont{Hirth}} \bibnamefont{and}
  \bibinfo{author}{\bibfnamefont{J.}~\bibnamefont{Lothe}},
  \emph{\bibinfo{title}{{Theory of Dislocations}}} (\bibinfo{publisher}{Wiley},
  \bibinfo{address}{New York}, \bibinfo{year}{1968}).

\bibitem[{\citenamefont{Weertman}(1955)}]{Weertman55}
\bibinfo{author}{\bibfnamefont{J.}~\bibnamefont{Weertman}},
  \bibinfo{journal}{Journal of Applied Physics} \textbf{\bibinfo{volume}{26}},
  \bibinfo{pages}{1213} (\bibinfo{year}{1955}).

\bibitem[{\citenamefont{Weertman and Weertman}(1975)}]{Weertman75}
\bibinfo{author}{\bibfnamefont{J.}~\bibnamefont{Weertman}} \bibnamefont{and}
  \bibinfo{author}{\bibfnamefont{J.~R.} \bibnamefont{Weertman}},
  \bibinfo{journal}{Annual review of earth and planetary sciences}
  \textbf{\bibinfo{volume}{3}}, \bibinfo{pages}{293} (\bibinfo{year}{1975}).

\bibitem[{\citenamefont{Frost and Ashby}(1982)}]{Frost82}
\bibinfo{author}{\bibfnamefont{H.~J.} \bibnamefont{Frost}} \bibnamefont{and}
  \bibinfo{author}{\bibfnamefont{M.~F.} \bibnamefont{Ashby}},
  \emph{\bibinfo{title}{{Deformation-Mechanism Maps: The Plasticity and Creep
  of Metals and Ceramics}}} (\bibinfo{publisher}{Pergamon Press},
  \bibinfo{address}{Oxford}, \bibinfo{year}{1982}).

\bibitem[{\citenamefont{{Van}~der Giessen and Needleman}(1995)}]{Giessen95}
\bibinfo{author}{\bibfnamefont{E.}~\bibnamefont{{Van}~der Giessen}}
  \bibnamefont{and}
  \bibinfo{author}{\bibfnamefont{A.}~\bibnamefont{Needleman}},
  \bibinfo{journal}{Model. Simul. Mater. Sci. Eng.}
  \textbf{\bibinfo{volume}{3}}, \bibinfo{pages}{689} (\bibinfo{year}{1995}).

\bibitem[{\citenamefont{Benzerga et~al.}(2004)\citenamefont{Benzerga,
  {Br\'echet}, Needleman, and Van~der Giessen}}]{Benzerga04MSMSE}
\bibinfo{author}{\bibfnamefont{A.~A.} \bibnamefont{Benzerga}},
  \bibinfo{author}{\bibfnamefont{Y.}~\bibnamefont{{Br\'echet}}},
  \bibinfo{author}{\bibfnamefont{A.}~\bibnamefont{Needleman}},
  \bibnamefont{and} \bibinfo{author}{\bibfnamefont{E.}~\bibnamefont{Van~der
  Giessen}}, \bibinfo{journal}{Model. Simul. Mater. Sci. Eng.}
  \textbf{\bibinfo{volume}{12}}, \bibinfo{pages}{159} (\bibinfo{year}{2004}).

\bibitem[{\citenamefont{Mordehai et~al.}(2008)\citenamefont{Mordehai, Clouet,
  Fivel, and Verdier}}]{Mordehai08}
\bibinfo{author}{\bibfnamefont{D.}~\bibnamefont{Mordehai}},
  \bibinfo{author}{\bibfnamefont{E.}~\bibnamefont{Clouet}},
  \bibinfo{author}{\bibfnamefont{M.}~\bibnamefont{Fivel}}, \bibnamefont{and}
  \bibinfo{author}{\bibfnamefont{M.}~\bibnamefont{Verdier}},
  \bibinfo{journal}{Phil. Mag.} \textbf{\bibinfo{volume}{88}},
  \bibinfo{pages}{899} (\bibinfo{year}{2008}).

\bibitem[{\citenamefont{Freund and Heinloth}(2002)}]{Freund02}
\bibinfo{author}{\bibfnamefont{P.}~\bibnamefont{Freund}} \bibnamefont{and}
  \bibinfo{author}{\bibfnamefont{K.}~\bibnamefont{Heinloth}},
  \emph{\bibinfo{title}{Numerical Data and Functional Relationships in Science
  and Technology: New Series}} (\bibinfo{publisher}{Springer},
  \bibinfo{address}{New York}, \bibinfo{year}{2002}).

\bibitem[{\citenamefont{Poirier}(1976)}]{Poirier76}
\bibinfo{author}{\bibfnamefont{J.~P.} \bibnamefont{Poirier}},
  \bibinfo{journal}{Rev. Phys. App.} \textbf{\bibinfo{volume}{11}},
  \bibinfo{pages}{731} (\bibinfo{year}{1976}).

\end{thebibliography}

\end{document}